\documentclass[nofootinbib,superscriptaddress,10pt, twocolumn]{revtex4-2}

\usepackage{bm}
\usepackage{latexsym}
\usepackage{dcolumn}
\usepackage{amsmath,amsfonts,amssymb}
\usepackage{graphicx,epsfig}
\usepackage{amsmath}
\usepackage{hyperref}
\usepackage{graphicx,epstopdf}
\usepackage{physics}
\usepackage[utf8]{inputenc}

\hypersetup{
	colorlinks   = true, 
	urlcolor     = blue, 
	linkcolor    = blue, 
	citecolor   = red 
}

\def\be{\begin{equation}}
	\def\ee{\end{equation}}
\def\beq{\begin{eqnarray}}
	\def\eeq{\end{eqnarray}}

\newcommand{\levicivita}{}
\def\levicivita#1#{\tensor#1{\epsilon}}

\newcommand{\Trf}[1]{\text{Tr}_{f}}

\begin{document}
	\title{Thermality of horizon through near horizon instability: a path integral approach}
		\author{Gaurang Ramakant Kane}
		\email{gsrkane@gmail.com}
		\affiliation{Department of Physics, Indian Institute of Technology Guwahati, Guwahati 781039, Assam, India}
	\affiliation{Rudolf Peierls Centre for Theoretical Physics, University of Oxford, Oxford OX1 3PU, United Kingdom}
	\author{Bibhas Ranjan Majhi}
	\email{bibhas.majhi@iitg.ac.in (Corresponding author)}
		\affiliation{Department of Physics, Indian Institute of Technology Guwahati, Guwahati 781039, Assam, India}

	\date{\today}
	
\begin{abstract}
Recent investigations revealed that the near horizon Hamiltonian of a massless, chargeless outgoing particle, for its particular motion in static as well as stationary black holes, is effectively $\sim xp$ kind. This is unstable by nature and has the potential to explain a few interesting physical phenomena. From the path integral kernel, we first calculate the density of states. Also, following the idea of [Phys. Rev. D 85, 025011 (2012)] here, in the vicinity of the horizon, we calculate the effective path corresponding to its Schrodinger version of Hamiltonian through the path integral approach. The latter result appears to be complex in nature and carries the information of escaping the probability of the particle through the horizon. In both ways, we identify the correct expression of Hawking temperature. Moreover, here we successfully extend the complex path approach to a more general black hole like Kerr spacetime. We feel that such a complex path is an outcome of the nature of near horizon instability provided by the horizon and, therefore, once again bolstered the fact that the thermalization mechanism of the horizon may be explained through the aforesaid local instability.
\end{abstract}

	
	\maketitle
	
	
\section{\label{Intro}Introduction and basic formalism}
A combination of Bekenstein's proposal \cite{Bekenstein:1973ur,Bekenstein:1974ax} and Hawking's explicit calculation \cite{Hawking:1974rv,Hawking:1975vcx} proposed black holes as thermodynamical objects. The temperature of the horizon is given by $T = (\hbar\kappa/2\pi)$ (in units of $c=1=k_B$), where $\kappa$ is known as surface gravity. Till now various approaches (e.g. path integral of gravitational action \cite{Gibbons:1976ue,Hawking:1976ja}, tunneling formalism \cite{Srinivasan:1998ty,Parikh:1999mf,Banerjee:2008cf,Banerjee:2009wb}, gravitational anomaly approach \cite{Robinson:2005pd,Iso:2006wa,Banerjee:2007qs,Banerjee:2008sn}, extremization of action functional for extremal black holes \cite{Sen:2005wa,Sen:2007qy,Majhi:2015pra}, Noether conserved charge \cite{Wald:1993nt,Iyer:1995kg,Strominger:1996sh,Carlip:1998wz,Majhi:2011ws}, etc.) have been put forward to get better understanding of the Hawking radiation phenomenon as well as thermal behaviour of  black holes. One of them was to visualise the radiation of the particle as the escaping of horizon barrier through a complex path -- known as quantum tunneling of particles \cite{Srinivasan:1998ty,Parikh:1999mf,Banerjee:2008cf,Banerjee:2009wb}. In this approach, the wave function of the particle is chosen in the semi-classical limit (the WKB approximation) as $\Psi \sim\exp[(i/\hbar)S_{HJ}]$, where $S_{HJ}$ is the Hamilton-Jacobi action. It has been observed that $S_{HJ}$ picks up a complex part across the horizon, providing a quantum mechanical probability for the particle to escape from the black hole, leading to a Boltzmann-like tunneling probability. An analogy with the usual Boltzmann expression yielded an exact expression for Hawking temperature. Interestingly, the choice of the wave function under the WKB approximation has a close similarity with the path integral form $\bra{x_2,t_2}\ket{x_1,t_1} = \mathcal{N}\int Dx \exp[(i/\hbar) S] \sim \mathcal{N}' \exp[(i/\hbar)S]$ in the semi-classical limit along with saddle point approximation \cite{Book1}, wherein the left hand side the ket and bra are basis vectors in Heisenberg picture. Here $S$ is the classical action for the system. In the Schrodinger picture, this path integral can be interpreted as the wave function after time $(t_2-t_1)$ in position representation which was initially located at $x_1$ at time $t_1$. Therefore evaluation of it under the aforementioned approximations must provide the wave function under WKB approximation, used in tunneling formalism. Hence obtention of the imaginary part of the action along the horizon provides a non-trivial signature about the black hole both in WKB or path integration approaches. 

The above discussion implies that the emission of the particle can occur only when the action incorporates an imaginary part in it. Hence in principle, the variational principle of this action must provide a complex path at the quantum level for the occurrence of such an event. In original tunneling formalism, no one has addressed the structure of the path. Interestingly, one can address such questions in the path integral approach to quantum mechanics. We already mentioned that at the semi-classical limit, both WKB wave function and path integration carry similar information. Therefore one expects that path integration must have a major role in providing a better understanding of this issue. The idea is as follows. In the presence of source $J(x,t)$, the vacuum to vacuum transition amplitude (VVTA) can be expressed in terms of path integration \cite{Book1}:
\begin{equation}
\bra{0^+}\ket{0^-}_J = \lim \limits_{\substack{%
t_1\to -\infty\\
t_2\to\infty}}\mathcal{N}\int Dx e^{\frac{i}{\hbar}S[x,J]} =
\lim \limits_{\substack{%
t_1\to -\infty\\
t_2\to\infty}}  Z[J]~.
\label{1}
\end{equation}
In this regard the effective action $W[J]$ is defined by $W[J] = -i\hbar\ln Z[J]$. Then VVTA can be expressed as 
\begin{equation}
\bra{0^+}\ket{0^-}_J = \lim \limits_{\substack{%
t_1\to -\infty\\
t_2\to\infty}} e^{\frac{i}{\hbar}W[J]}~,
\label{2}
\end{equation}
and hence the transition probability is given by the imaginary part of the effective action:
\begin{equation}
\Big|\bra{0^+}\ket{0^-}_J\Big|^2 = \lim \limits_{\substack{%
t_1\to -\infty\\
t_2\to\infty}} e^{-\frac{2W_{I}[J]}{\hbar}}~.
\label{3}
\end{equation}
In the above, $W_I$ is the imaginary part of the effective action $W$. The above discussion implies that the probability other than unity occurs only when $W_I$ is non-vanishing. In that situation, the particle jumps to other states with finite (non-trivial) probability. Therefore the transition of a quantum mechanical system happens due to source $J$ only if the effective action $W$ is complex. 
Applying this theory to a forced quantum oscillator by a time-dependent source produces the correct transition probability value (see a detailed discussion in \cite{Book1,ComplexPath1}). In fact, the energy transferred ($\mathcal{E}_0$) by the source is shown to be related to the transition amplitude and hence can be determined from the imaginary part of the effective action ($\mathcal{E}_0 = W_I$). This idea has also been introduced in quantum field theory, and it is found that the imaginary part of the effective action clearly explains the particle production (one of the well-known examples is the Schwinger effect \cite{Book2}). Therefore a general consciousness is -- the appearance of complex effective action is the signature of transition from the ground state to other states in quantum mechanics and particle production in quantum field theory.

 In this regard, the effective path can be obtained by extremising the aforesaid effective action $W$. In fact, the average path is determined by \cite{ComplexPath2}
\begin{equation}
X(t) = \mathcal{N}\int Dx ~ x(t) e^{\frac{i}{\hbar}S[x,J]}~,
\label{4}
\end{equation}
and therefore alternatively we have
\begin{eqnarray}
X(t) &=& -i\hbar \frac{1}{Z[J]}\frac{\delta Z[J]}{\delta J(t)}\Big|_{J=0} = \frac{\delta W[J]}{\delta J(t)}\Big|_{J=0} 
\nonumber
\\
&=& \frac{\bra{x_2,t_2}x(t)\ket{x_1,t_1}}{\bra{x_2,t_2}\ket{x_1,t_1}}~.
\label{5}
\end{eqnarray}
In the above, like earlier, $J$ acts as an external source and therefore $S[x,J]$ is expressed as $S[x,J] = S[x] + \int dt J(t)x(t)$. Here $S[x]$ is the classical action for the system under consideration. Hence (\ref{5}) can be applied to any quantum mechanical system.
Hence the signature of complex effective action must be reflected in the path $X(t)$ as well, and so the complex nature of the path is the confirmation of the transition from the ground state to other states. The conception of the path through the above definition was initially put forward in \cite{ComplexPath2} but did not get attention until its application was made in \cite{ComplexPath1}. The above form can be cast in terms of propagators as well:
\begin{equation}
X(t)=\dfrac{\int_{-\infty}^{\infty} dx~x(t)K(x_{2},t_{2};x,t)K(x,t;x_{1},t_{1})}{K(x_{2},t_{2};x_{1},t_{1})}~.
\label{6}
\end{equation}
This has been extensively used in \cite{ComplexPath1}. In fact in \cite{ComplexPath1}, the authors showed for forced harmonic oscillator that the imaginary part of $W$ is proportional to $|X(t)|^2$ in the limit $t\to\infty$ and so $|X(t)|^2$ is directly connected to the transition probability (see Eq. (\ref{3}) for the general relation between transition probability and $W_I$). Such an example lets them conjecture that {\it the probability transition or the particle production can be well studied through the complex nature of the path $X(t)$ as well}. In fact, as the energy transferred by the source is given by $W_I$, they drew an analogy through the results of the forced harmonic oscillator that $d|X(t)|^2/dt$ provides information about the rate of energy transferred.  

As an application of the conjecture, the authors of \cite{ComplexPath1} calculated $d|X(t)|^2/dt$ for the effective potential as seen by the scalar modes in the near horizon limit. The evaluation of $X(t)$ just around the horizon shows a complex nature. More interestingly, $d|X(t)|^2/dt$ appears to be of the form given by the energy spectrum of a quantum harmonic oscillator  $\sim (N+1/2)$. For harmonic oscillator $N$ denotes $N^{th}$ quantum state and when it is in this state the total energy is given by $(N+1/2)$ times the energy of a ``photon''. Therefore $N$ can be treated as the number of emitted ``photons'' contained within emitted energy when the excited harmonic oscillator decays to ground state. With this analogy the value of $N$, calculated from complex path can be treated as the number of the radiated particles due to the energy transferred by the potential. Interestingly in this case $N$ is given by the Planckian distribution. This led to the identification of the horizon temperature. Apart from the conjecture, a few interesting observations came in the study \cite{ComplexPath1}, which we want to mention below.
\begin{itemize}
\item The temperature was found to suffer from a factor of $2$ discrepancy. It has been argued in the paper that such is due to the choice of Schwarzschild-like coordinates. Moreover, the analysis was done only for static spherically symmetric black holes (SSSBH).
\item An important comment has been made in the conclusion of \cite{ComplexPath1} that a necessary (but not sufficient) condition for obtaining a complex path in this procedure is the Hamiltonian must be unbounded and/or non-hermitian (actually, the Hamiltonian needs to be non-self-adjoint operator, rather than non-hermitian \cite{Private1}).
\end{itemize}
In the present article, we want to follow up on the above idea and see how far the idea can be extended. We aim to extend the idea of the complex path beyond SSSBH (e.g. Kerr black hole) and also see whether a correct temperature value can be obtained. 

Recently one of the authors of this paper has been trying to understand the thermal nature of horizons using a very novel idea based on a model consisting of a chargeless massless particle moving very near the horizon. The model shows that the near horizon Hamiltonian of the particle is effectively given by the following form
\begin{equation}
H=\kappa xp~,
\label{eqn:near_horizon_hamiltonian}
\end{equation}
where $x$ is the radial distance from the horizon and $p$ is its conjugate momentum. So $x=0$ corresponds to the location of the horizon. In the above, $\kappa$ is the surface gravity. Such a Hamiltonian has been obtained in Eddington- Finkelstein (EF) outgoing null coordinates both for a general SSSBH as well as Kerr spacetime. In the above, the classical path of the particle has been chosen to be along the normal to null hypersurface, given by Eddington null coordinate $u =$ constant, where $u = t - r_*$ with $r_*$ is the well known tortoise coordinate (see \cite{Chaos3,Dalui:2021tvy} for details regarding the construction of (\ref{eqn:near_horizon_hamiltonian})). The same can also be obtained in Painleve coordinates for SSSBH as well as Kerr BH, considering a particular trajectory \cite{Dalui:2018qqv,chaos1,chaos2} (a generic null surface also provides such Hamiltonian \cite{Dalui:2021sme}). We just mentioned that the form of the Hamiltonian (\ref{eqn:near_horizon_hamiltonian}) was obtained explicitly in two specific choices of coordinates -- one in EF and another in Painleve coordinates. Therefore the time can be taken at this stage as either EF time or Painleve time. Moreover, it must be explicitly mentioned that such a Hamiltonian was obtained for a specific choice of outgoing null path. For instance in EF coordinates the tangent of the path is normal to $u=$ constant hypersurface, while in Painleve we found this for radially outgoing null path. In this sense the Hamiltonian is very specific to these choices of null paths, and also it depends on a particular choice of observer. We will see that the Hawking temperature can be obtained by using the Hamiltonian. Since it is well known that horizon thermodynamics is an observer dependent concept, we can expect the underlying dynamics may depend on choice of coordinates. However it would be interesting to investigate the generality of such form of Hamiltonian. 
Also note that here $x = 0$ is location of the horizon. As the Hamiltonian has been obtained by expanding the metric coefficients around $x=0$ and keeping only the leading order term, the structure of the Hamiltonian retains for both $x>0$ (just outside the horizon) and $x<0$ (just inside the horizon). Therefore the applicability of the Hamiltonian is valid in the range $-\epsilon\leq x\leq \epsilon$ with $\epsilon>0$, where $\epsilon$ is a very small quantity. We will use this in our latter analysis.

Nonetheless, it must be noted that the above Hamiltonian, very near the horizon, is unstable. First discuss for $x>0$. The solutions of the equations of motions are given by $x\sim e^{\kappa t}$ and $p\sim e^{- \kappa t}$. Since the near horizon limit $x\to 0$ is equivalent to $t\to -\infty$, the momentum diverges there. Moreover for a fixed energy (\ref{eqn:near_horizon_hamiltonian}) implies that $p\to\infty$ as $x\to 0$. Both imply that the Hamiltonian has radial instability in the near horizon regime. On the other hand, (\ref{eqn:near_horizon_hamiltonian}) can be cast to that of an inverted harmonic oscillator (see \cite{chaos2}), which is unstable in nature. While for $x<0$, the momentum is given by $p = -(H/\kappa y)$, where $x=-y$ with $y>0$. This again diverges at the horizon $y=0$. In this case the classical solutions take the forms as $y\sim e^{-\kappa t}$ and $p\sim e^{\kappa t}$. Since here $y\to 0$ implies $t\to \infty$, one observes that the classical value of $p$ also diverges at the horizon.
Recently we explicitly showed this “local instability” may cause the temperature to the horizon at the semi-classical level \cite{chaos2,Chaos3,Dalui:2021tvy,Chaos4}.

Inspired by all these facts here, we like to investigate whether the unstable Hamiltonian (\ref{eqn:near_horizon_hamiltonian}) can show a complex path and thereby produce a correct form of Hawking temperature. We will show by constructing an effective non-relativistic Hamiltonian in the Schrodinger description that this is indeed the case. Since this Hamiltonian is related to both Kerr and SSSBH spacetimes, our model is capable of presenting a very general situation, rather than restricted to SSSBH as was done originally in \cite{ComplexPath1}. Moreover, we will see that the identified temperature does not suffer from the factor of two ambiguity. This shows that our choice of coordinates in which (\ref{eqn:near_horizon_hamiltonian}) has been obtained are suitable ones. The same is also being confirmed by obtaining the density of states directly from original relativistic $\sim xp$ Hamiltonian. Additionally, as (\ref{eqn:near_horizon_hamiltonian}) is unstable (or unbounded) in nature, the outcome from this again confirms the robustness of the corollary, suggested in \cite{ComplexPath1} -- {\it the necessary (but maybe not sufficient) condition that the Hamiltonian must be unbounded in order to obtain a complex path}.

Let us now move forward toward the calculation in favour of our claim. In this analysis the Hamiltonian (\ref{eqn:near_horizon_hamiltonian}) will be treated as that of a quantum mechanical system and calculation will be done using the standard prescriptions of path integral formalism. Before this, we will first calculate the density of states corresponding to (\ref{eqn:near_horizon_hamiltonian}) using the path integral kernel and show that such is thermal in nature. Later an effective Hamiltonian will be constructed under the Schrodinger description in which case the path will be calculated.


\section{Density of states and the ground state}\label{DOS}
The density of states (DOS) corresponding to our present model can be evaluated from the kernel or propagator $K(x_2,t_2;x_1,t_1)$. In path integral approach DOS is given by \cite{Book3,Book4}
\begin{equation}
\rho(E) = -\frac{1}{\pi} \textrm{Im}\Big(\int G(X,E)dX\Big)~,
\label{7}
\end{equation}
where $G$ is the Laplace-Fourier transformation of $K$:
\begin{equation}
G(X,E) = \int_0^\infty dt~e^{-\frac{iEt}{\hbar}} K(x_2=X,t_2=t;x_1=X,t_1=0)~.
\label{8}
\end{equation}
The propagator for the Hamiltonian (\ref{eqn:near_horizon_hamiltonian}) is given by \cite{Chaos4}
\begin{equation}
K(x_2,t_2=t;x_1,t_1=0) = e^{-\frac{\kappa t}{2}}\delta\Big(x_1 - x_2 e^{-\kappa t}\Big)~,
\label{7a}
\end{equation}
with $t>0$. 
\begin{widetext}
Then we find
\begin{eqnarray}
\int_{-\infty}^{\infty} dX
G(X,E) &=& \int_{-\infty}^{+\infty}dX\int_{0}^{+\infty}dt e^{-\frac{iE t}{\hbar}} K(x_2=X,t_2=t;x_1=X,t_1=0)
\nonumber
\\
&=& \int_{-\infty}^{+\infty}dX\int_{0}^{+\infty}dt e^{-\frac{iE t}{\hbar}}  e^{-\frac{\kappa t}{2}}\delta\Big(X - X e^{-\kappa t}\Big)~.
\label{10a}
\end{eqnarray}
The above integrant for $t$-integration has poles where the argument of the Dirac-Delta function vanishes. The $t$-integration can be performed by the complex integration method. In order to do that, we first replace $t\to t-i\delta$ with $\delta>0$, and after completing the integration, the limit $\delta\to 0$ will be taken. Then the poles are given by $e^{\kappa (t-i\delta)} = 1$; i.e. $t=i\delta\pm(2\pi i n)/\kappa$ with $n=0,1,2,3, \dots$. Now for $E>0$ in (\ref{10a}) only the poles on the lower imaginary axis will contribute; i.e. we have now $t= - (2\pi i n)/\kappa$ with $n=1,2,3,\dots$. In that case, the contributing part of the Dirac-Delta function can be expressed as
\begin{eqnarray}
\delta\Big(X - X e^{-\kappa t}\Big) &=& \sum_{n=1,2,3,\dots}\frac{\delta\Big(t+\frac{2\pi i n}{\kappa}\Big)}{\frac{d}{dt}(X-Xe^{-\kappa t})\Big|_{t=-\frac{2\pi i n}{\kappa}}}
\nonumber
\\
&=& \frac{1}{\kappa X}\sum_{n=1,2,3,\dots} \delta\Big(t+\frac{2\pi i n}{\kappa}\Big)~.
\label{11a}
\end{eqnarray}
Substituting this in (\ref{10a}) and performing the integration one finds
\begin{eqnarray}
\int_{-\infty}^{\infty} dX
G(X,E) &=& \int_{-\infty}^{\infty} \frac{dX}{\kappa X}\sum_{n=1,2,3,\dots} \Big(-e^{-\frac{2\pi E}{\hbar\kappa}}\Big)^n = -\frac{1}{\kappa}\frac{1}{e^{\frac{2\pi E}{\hbar\kappa}}+1} \int_{-\infty}^{\infty} \frac{dX}{X}~
\nonumber
\\
&=& -\frac{i\pi}{\kappa}\frac{1}{e^{\frac{2\pi E}{\hbar\kappa}}+1} + \textrm{Real part}~.
\label{12a}
\end{eqnarray}
In the step we have used the relation $ \int_{-\infty}^{\infty} \frac{dX}{X} = i\pi + \textrm{P} \Big( \int_{-\infty}^{\infty} \frac{dX}{X}\Big)$, where ``P'' stands for the principal value.
\end{widetext}
Therefore DOF states, by Eq. (\ref{7}), is given by
\begin{equation}
\rho(E) =  \frac{1}{\kappa}\frac{1}{e^{\frac{2\pi E}{\hbar\kappa}}+1}~.
\label{12}
\end{equation}
The same can be obtained by transforming the Hamiltonian (\ref{eqn:near_horizon_hamiltonian}) in the form of that of an inverted harmonic oscillator (see Appendix \ref{App3}).
Note that it satisfies the Kubo's form of fluctuation-dissipation relation \cite{Kubo}
\begin{equation}
\rho^+(E) = \coth(\frac{\beta E}{2})\rho^-(E)~,
\label{13}
\end{equation}
if one identifies the inverse temperature as $\beta = 2\pi/\hbar\kappa = 1/T$, where $\rho^{\pm}(E) = \rho(-E)\pm\rho(E)$. Note that $\beta$ is the inverse Hawking temperature.

It may be noted that (\ref{12}) refers to the number of modes at energy $E$ which are at temperature $T$. If these are Hawking radiated ones, then the corresponding entropy can be interpreted as that of the Hawking radiation. In principle this can be calculated following works of Page \cite{Page1,Page2,Page3}. By integrating (\ref{12}) over all possible energies, the energy emission rate $dE/dt$ can be obtained. Then introducing the law of thermodynamics $TdS/dt = dE/dt$ one finally calculates the rate of entropy change of the radiation.

Let us now investigate another aspect of the Hamiltonian (\ref{eqn:near_horizon_hamiltonian}) through path integral or, in other words, through the propagator. It is well formulated that 
in Euclidean time $t=-it_E$ formalism, the ground state energy can be determined from the propagator. It is given by
\begin{equation}
E_0 = \lim_{t_E \to \infty} \Big[-\frac{\hbar}{t_E}\ln \Big(\frac{\bra{t_E,\bf{0}}\ket{0,\bf{0}}}{|\phi_0({\bf{0}})|^2}\Big)\Big]~.
\label{E1}
\end{equation}
where $\bra{t_E,\bf{0}}\ket{0,\bf{0}} = G_E(t_E,{\bf 0};0,{\bf 0})$. Here $\phi_0({\bf x})$ is the ground state wave function, can be evaluated by the following relation:
\begin{equation}
\phi_0({\bf x}) = N \lim_{t_E\to\infty} K_E(t_E\to \infty,{\bf 0}; 0, {\bf x})~,
\label{E2}
\end{equation}
with $N$ is the normalization factor. These definitions lead to the required expressions when one retains only the leading order terms in the limit $t_E\to \infty$ (for a detailed discussion, see section $1.2.3$ of \cite{Book2}).
For the present case, the propagator is given by (\ref{7a}). As it already has the damping property with respect to $t$, we will use the above Euclidean time formalism by taking $t_E=t$. Under this assumption setting $x_1=x$ and $x_2=0$ in (\ref{7a}) and using (\ref{E2}) we find \begin{equation}
\phi_0(x) = N \lim_{t\to\infty} e^{-\frac{\kappa t}{2}}\delta(x)~.
\label{E3}
\end{equation}
It may be noted that at $t\to\infty$, $\phi_0(0)$ can be finite as $e^{-\kappa t/2}$ is a decaying function while $\delta(0)$ is a diverging one. Therefore in order to calculate (\ref{E1}) we substitute $\phi_0(0)$ without taking the limit $t\to\infty$ at this stage. But this will be taken at the final stage, as mentioned in (\ref{E1}) for the definition of $E_0$.
With this the ground state energy is calculated to be
\begin{eqnarray}
E_0 &=& -\lim_{t\to\infty}\frac{\hbar}{t} \ln\Big[\frac{e^{-\frac{\kappa t}{2}}\delta(0)}{N^2\delta(0)\delta(0)e^{-\kappa t}}\Big]
\nonumber
\\
&=& -\frac{\hbar\kappa}{2}+\lim_{t\to\infty}\frac{\hbar}{t}\ln\Big[N^2\delta(0)\Big]~.
\label{E4}
\end{eqnarray} 
The above one seems to be divergent. The renormalised one can be evaluated by the following argument. In the absence of the acceleration (i.e. choose an observer, e.g. a freely falling observer, for which acceleration vanishes), the contribution to $E_0$ is given by
\begin{equation}
E_0^{(\kappa=0)} = \lim_{t\to\infty}\frac{\hbar}{t}\ln\Big[N^2\delta(0)\Big]~.
\label{E5}
\end{equation}
This amount can be treated as the background contribution. Subtracting this amount from (\ref{E4}) one obtains the renormalised value of ground state energy of the particle as
\begin{equation}
E_0^{(ren)} = E_0^{\kappa=0} - E_0 =\frac{\hbar\kappa}{2}~.
\label{E6}
\end{equation}

Few observations and comments are as follows.
\begin{enumerate}
\item The renormalized ground state energy (\ref{E6}) is similar in form to that of one dimensional quantum Harmonic oscillator ($E_0^{(HO)} = (1/2)\hbar\omega$).
\item Being this as the ground state energy, it can be treated as the minimum energy of the particle. Then by (\ref{eqn:near_horizon_hamiltonian}) we have $(xp)_{\textrm{min.}} = \hbar/2$, which is quite consistent with the minimum uncertainty relation.
\item The ground state energy can be expressed in terms of Hawking temperature as 
\begin{equation}
E_0^{(ren)} = \pi T~.
\label{E8}
\end{equation} 
Considering a thermal equilibrium between the particle and the horizon, one finds that $T$ is the temperature of the particle as well. This implies that the ground state (or minimum) energy is purely thermal in nature.
\end{enumerate}
Here, in order to obtain a finite value of the ground state energy (Eq. (\ref{E6})) we have subtracted the divergent part. This makes the result finite and meaningful. This type of approach has been used earlier in various situations where one removes the \emph{background contribution} to get a finite result. But, this argument is speculative and needs further investigation to see the correctness of the argument.

\section{An effective quantum model}
In this later discussion our aim is two fold. Particularly we are interested to investigate whether a non-relativistic potential, which corresponds to the eigenstates of original Hamiltonian (\ref{eqn:near_horizon_hamiltonian}) through Schrodinger equation, also carries thermality. Additionally we like to study whether a quantum path can be obtained which provides information about the horizon temperature. In this regard it must be pointed out that the Hamiltonian, which will be obtained here through a non-relativistic treatment, is not equivalent to original Hamiltonian (\ref{eqn:near_horizon_hamiltonian}). This is because we will see that the propagators for these two Hamiltonians do not match; only they provide same eigenstates in the near horizon limit. But as far as our main objective is concerned, we are interested to look into their role in the explanation of horizon thermality. In achieving the path, it must be mentioned that in literature the analysis for the same is given 
for a non-relativistic Hamiltonian of the form:
\begin{equation}
	H=\dfrac{p^{2}}{2m}+V(x)
	\label{eqn:standard_hamiltonian}~~,
\end{equation}
where, $p$ is the momentum operator and $V(x)$ is the potential. The present Hamiltonian (\ref{eqn:near_horizon_hamiltonian}) is not in this form. In the position space, this Hamiltonian gives a first order differential equation. Furthermore, the path integral formalism in quantum mechanical systems is generally used for Hamiltonians of the form given in Eq. \eqref{eqn:standard_hamiltonian}. Therefore, here, we first investigate which Hamiltonian of the form Eq. \eqref{eqn:standard_hamiltonian} has the wave function solution same as that of the Hamiltonian presented in (\ref{eqn:near_horizon_hamiltonian}). We will see that the Hamiltonian has a $1/x^{2}$ type potential {\footnote{This form of potential, related to near horizon of a black hole spacetime, has also been obtained and studied in \cite{Camblong:2020pme,Azizi:2020gff} (see also \cite{Camblong:2022oet}).}}. Since this not the same as (\ref{eqn:near_horizon_hamiltonian}), we call the present model as an effective quantum model (not the original one) and therefore the potential is termed as non-relativistic effective potential of our actual model.

To move towards the quantum mechanics let us first make the classical Hamiltonian (\ref{eqn:near_horizon_hamiltonian}) as a hermitian operator $\hat{H} = \dfrac{\kappa}{2}(\hat{x}\hat{p}+\hat{p}\hat{x})$. Then the time-dependent evolution equation for the state of the particle is given by
\begin{equation}
	i\hbar\dfrac{d\Psi}{dt}=H\Psi=\dfrac{\kappa}{2}(\hat{x}\hat{p}+\hat{p}\hat{x})\Psi~.
	\label{eqn:schrodinger}
\end{equation}
As the Hamiltonian is time-independent, we can write $\Psi(x,t)=\psi(x)e^{-\frac{iEt}{\hbar}}$. Here, $E$ is the energy of the massless particle. This yields the time-independent evolution equation in position representation as 
\begin{equation}
-\frac{i\hbar\kappa}{2}\Big(2x\frac{d\psi}{dx}+\psi\Big) = E\psi~.
\label{eqn:schrodinger1}
\end{equation}
In order to provide the above equation in the form of standard Schrodinger equation, differentiate the above one once. Then after a trivial rearrangement we obtain
\begin{equation}
	-\dfrac{\hbar^{2}}{2m}\dfrac{d^{2}\psi}{dx^{2}}+\dfrac{\hbar^{2}}{8mx^{2}}\Big(-\dfrac{8iE}{\hbar\kappa}+3-\dfrac{4E^{2}}{\hbar^{2}\kappa^{2}}\Big)\psi=0~.
\end{equation}
In the above a parameter $m$ has been introduced in order to give the above equation as a time-independent Schrodinger equation form
\begin{equation}
	\Big[\dfrac{\hat{p}^{2}}{2m}+V(x)\Big]\psi(x)=\mathcal{E}\psi~.
\label{B1}
\end{equation}
Note that our original relativistic near horizon Hamiltonian (\ref{eqn:near_horizon_hamiltonian}) now provides a non-relativistic effective evolution of the eigenstate through the Schrodinger equation of the form (\ref{B1}) under an effective potential of the following form
\begin{equation}
V(x)=-\dfrac{k}{x^{2}};~~~k=\dfrac{\hbar^{2}}{8m}\Big(\dfrac{8iE}{\hbar\kappa}-3+\dfrac{4E^{2}}{\hbar^{2}\kappa^{2}}\Big)~,
\label{eqn:potential}
\end{equation}
with the energy of the non-relativistic particle of mass $m$ is given by $\mathcal{E}=0$. The same can also be obtained using the solutions of the actual wave equation (\ref{eqn:schrodinger1}) (i.e. the energy eigenfunctions) and asking which potential is responsible for such eigenfunctions as solutions of the Schrodinger equation (see a discussion in Appendix \ref{App1}). Further, in Appendix \ref{App1} we solve the Schrodinger equation for $1/x^{2}$ potential and show that near $x=0$, the wavefunction will have a similar form for energies $\mathcal{E}\geq 0$. This is an exciting result as people have worked out the thermal behaviour of $1/x^{2}$ type potential \cite{ComplexPath1, ComplexPath2}.



\section{Complex Effective Path and temperature}
Now we proceed to calculate the effective path $X(t)$ of the system through the definition (\ref{6}). As mentioned in the introduction and \cite{ComplexPath1}, this quantity is complex in general and is related to vacuum to vacuum transition probability. In order to obtain this we need to first know the expression of the propagator $K(x_2,t_2; x_1,t_1)$.

The propagator (also called as the kernel) for a system with potential $V(x)=-k/x^{2}$ is given by \cite{ComplexPath1},
\begin{eqnarray}
&&K(x_{2},t_{2};x_{1},t_{1})=e^{-\frac{1}{2}i\pi(\gamma+1)}\Big(\dfrac{m}{2\hbar(t_{2}-t_{1})}\Big)(x_{1}x_{2})^{1/2}
\nonumber
\\
&\times&\exp\Big[\dfrac{im(x_{1}^{2}+x_{2}^{2})}{2\hbar(t_{2}-t_{1})}\Big]H_{\gamma}^{(2)}\Big(\dfrac{mx_{1}x_{2}}{\hbar(t_{2}-t_{1})}\Big)
\label{eqn:propagator}
\end{eqnarray}
where, $m$ is the mass of the particle and $(x_{1},t_{1})$, $(x_{2},t_{2})$ are the starting and the end points respectively. $H^{(2)}_{\gamma}$ is the Hankel function of second kind with,
\begin{equation}
\gamma=\sqrt{\dfrac{1}{4}-\dfrac{2mk}{\hbar^{2}}}~.
\label{eqn:definition_of_a}
\end{equation}

Further, we define $a=-i\gamma$. This is simply done for keeping our notation consistent with the \cite{ComplexPath1}. This will help us to use the results of this article directly. Note that the above propagator is different from (\ref{7a}) and therefore this model is completely different from our original relativistic Hamiltonian (\ref{eqn:near_horizon_hamiltonian}). But since our final interest to check whether a non-relativistic Hamiltonian whose energy eigenstates are also those of (\ref{eqn:near_horizon_hamiltonian}) can carry information about the thermal behavior of horizon, it is worth to investigate in a rigorous manner.

\begin{widetext}
Substituting \eqref{eqn:propagator} in (\ref{6}) one obtains
\begin{eqnarray}
X(t)&=&\lambda e^{-\frac{i\pi}{2}(ia+1)}\exp\Big[\dfrac{im}{2\hbar}\Big(\dfrac{x_{1}^{2}}{t-t_{1}}+\dfrac{x_{2}^{2}}{t_{2}-t}-\dfrac{(x_{1}^{2}+x_{2}^{2})}{t_{2}-t_{1}}\Big)\Big]
\Big[H_{ia}^{(2)}\Big(\dfrac{mx_{1}x_{2}}{\hbar(t_{2}-t_{1})}\Big)\Big]^{-1}
\nonumber
\\
&\times&\int_{-\infty}^{\infty}dx~x^{2}e^{i\lambda x^{2}} H_{ia}^{(2)}(px)H_{ia}^{(2)}(qx)
\label{effectivepathdefinition}
\end{eqnarray}
where, 
\begin{equation}
\lambda=\dfrac{m(t_{2}-t_{1})}{2\hbar(t_{2}-t)(t-t_{1})};~~p=\dfrac{mx_{1}}{\hbar(t-t_{1})};~~q=\dfrac{mx_{2}}{\hbar(t_{2}-t)}~.
\label{eqn:different_symbols}
\end{equation}
Our main goal is to see whether the above quantity provides any information about the escape of the particle through the horizon. Therefore, as argued in \cite{ComplexPath1}, it is sufficient to evaluate (\ref{eqn:different_symbols}) just around $x=0$. In the previous sections, we also saw that the $1/x^{2}$-type model is conidered near $x=0$, where the wave function solutions of the two different forms of Hamiltonian are the same. Along with this, we need to incorporate the effects of singularity at $x=0$, effectively leading to transition probability. To study the system near $x=0$ we choose $x_{1}=-\epsilon, x_{2}=+\epsilon, t_{1}=0, t_{2}\rightarrow \infty$ with $\epsilon\rightarrow 0^{+}$. The same was also incorporated in \cite{ComplexPath1}. The idea behind this as follows. Since we are interested in particle production, the particle should escape from the black hole horizon $x=0$. Usually it is conjectured that the Hawking radiation is a near horizon effect and the radiated particles quantum mechanically tunnel through the horizon. Hence in this picture the radiation happens when a particle starts just behind the horizon and escapes outside it. Since the path is from $x_1$ to $x_2$, therefore it is legitimate to choose $x_1 = -\epsilon$ and $x_2 = \epsilon$ with $\epsilon>0$. However at the end of the calculation we must take $\epsilon\to 0$ to mimic the near horizon nature of the phenomena.

On substituting the values of $x_{1},x_{2}$ in Eq. \eqref{effectivepathdefinition} we get,
\begin{equation}
	\begin{split}
		X(t)=\lambda e^{-\frac{i\pi}{2}(ia+1)}\exp\Big[\dfrac{im\epsilon^{2}}{2\hbar}\Big(\dfrac{1}{t-t_{1}}+\dfrac{1}{t_{2}-t}-\dfrac{2}{t_{2}-t_{1}}\Big)\Big]\Big[H_{ia}^{(2)}\Big(-\dfrac{m\epsilon^{2}}{\hbar(t_{2}-t_{1})}\Big)\Big]^{-1}\\\times \int_{-\infty}^{\infty}dx~x^{2}e^{i\lambda x^{2}} H_{ia}^{(2)}(px)H_{ia}^{(2)}(qx)~.
	\end{split}
	\label{eqn:effective_path}
\end{equation}
The limits, as mentioned above, will be taken at the end. To proceed further we denote
\begin{equation}
	I=\int_{-\infty}^{\infty}dx~x^{2}e^{i\lambda x^{2}} H_{ia}^{(2)}(px)H_{ia}^{(2)}(qx); ~~D=H_{ia}^{(2)}\Big(-\dfrac{m\epsilon^{2}}{\hbar(t_{2}-t_{1})}\Big)~.
	\label{eqn:numeratordenominator}
\end{equation}
Hence, in this notation the path is given by
\begin{equation}
	X(t)=-i\lambda e^{\frac{\pi a}{2}}\exp\Big[\dfrac{im\epsilon^{2}}{2\hbar}\Big(\dfrac{1}{t-t_{1}}+\dfrac{1}{t_{2}-t}-\dfrac{2}{t_{2}-t_{1}}\Big)\Big]\times\dfrac{I}{D}~.
\end{equation}

The main aim is now to evaluate the integral $I$. For the same, we expand the Hankel function in terms of Bessel Functions using the following relation:
\begin{equation}
	H_{\nu}(z)=\dfrac{J_{-\nu}(z)-e^{i\pi\nu}J_{\nu}(x)}{-i\sin(\nu\pi)}~.
\end{equation}
Substituting this, we get,
\begin{equation}
	I=I_{1}+I_{2}+I_{3}
\end{equation}
with,
\begin{eqnarray}
	I_{1}=\dfrac{e^{-2\pi a}}{\sinh^{2}(\pi a)}\int_{-\infty}^{\infty}dx~x^{2}e^{i\lambda x^{2}} J_{ia}(px)J_{ia}(qx)~;
	\label{B3}
	\\
	I_{2}=\dfrac{1}{\sinh^{2}(\pi a)}\int_{-\infty}^{\infty}dx~x^{2}e^{i\lambda x^{2}} J_{-ia}(px)J_{-ia}(qx)~;
	\label{B4}
	\\
	I_{3}=-\dfrac{2e^{-\pi a}}{\sinh^{2}(\pi a)}\int_{-\infty}^{\infty}dx~x^{2}e^{i\lambda x^{2}} (J_{-ia}(px)J_{ia}(qx)+J_{ia}(px)J_{-ia}(qx))~.
\label{B5}
\end{eqnarray}
In the limit, $\epsilon\rightarrow 0$, $D$ in Eq. \eqref{eqn:numeratordenominator} can be expanded as,
\begin{equation}
D=H^{(2)}_{ia}\Big(-\dfrac{m\epsilon^{2}}{\hbar(t_{2}-t_{1})}\Big)\approx \dfrac{i2^{-ia}\Gamma(-ia)}{\pi}\Big(\dfrac{m\epsilon^{2}}{\hbar(t_{2}-t_{1})}\Big)^{ia}+\dfrac{i2^{ia}\Gamma(ia)}{\pi}\Big(\dfrac{m\epsilon^{2}}{\hbar(t_{2}-t_{1})}\Big)^{-ia}e^{-\pi a}~.
\label{B6}
\end{equation}

Note that unlike in \cite{ComplexPath1}, in our case, the ``$a$" in Eq. \eqref{eqn:definition_of_a} is a complex value, given by,
\begin{equation}
	a=-a_{0}-i;~~~a_{0}=\dfrac{E}{\hbar\kappa}~.
\end{equation}
Substituting this value, in the limit $\epsilon\rightarrow 0$, we get (see Appendix \ref{App2}),
\begin{equation}
\dfrac{I_{1}+I_{2}}{D}=-e^{-\pi a_{0}/2}(1+e^{2\pi a_{0}})\dfrac{(-i\lambda)^{-3/2}}{2}\sqrt{\dfrac{a_{0}}{\cosh(\pi a_{0})\sinh(\pi a_{0})}}e^{i(\theta_{3}+\theta_{4})}
\label{B8}
\end{equation}
and,
\begin{equation}
\dfrac{I_{3}}{D}=0~.
\label{B9}
\end{equation}
The $\theta_i$s, which are real, are defined in Appendix \ref{App2}. 
Then Eq. \eqref{eqn:effective_path} reduces to,
\begin{eqnarray}
X(t)&=&i\lambda e^{\pi a/2}e^{-\pi a_{0}/2}(1+e^{-2\pi a_{0}})\dfrac{(-i\lambda)^{-3/2}}{2}\sqrt{\dfrac{a_{0}}{\cosh(\pi a_{0})\sinh(\pi a_{0})}}e^{i(\theta_{3}+\theta_{4})}
\nonumber
\\
&=& \lambda e^{-\pi a_{0}}(1+e^{2\pi a_{0}})\dfrac{(-i\lambda)^{-3/2}}{2}\sqrt{\dfrac{a_{0}}{\cosh(\pi a_{0})\sinh(\pi a_{0})}}e^{i(\theta_{3}+\theta_{4})}
\nonumber
\\
&=& \lambda(-i\lambda)^{-3/2}\sqrt{\dfrac{a_{0}\cosh(\pi a_{0})}{\sinh(\pi a_{0})}}e^{i(\theta_{3}+\theta_{4})}~.
\end{eqnarray}
Note that to achieve the above form we considered $x_1 = -\epsilon = -x_2$ and used $\epsilon\to 0$. These are very essential in order to reach at the required form which we will obtain in the subsequent analysis. The choice of $x_1,x_2$, as mentioned earlier, strictly from the physical aspects of the phenomena. While $\epsilon\to 0$ is essential to restrict the calculation in a very near horizon region which helps to simplify various mathematical steps, particularly see the evaluation of the integrations done in Appendix \ref{App2}.
\end{widetext}

It is already being discussed in the introduction that the vacuum to vacuum transition probability is given by the imaginary part of the effective action (see Eq. (\ref{3})). Therefore, as well known in literature, the complex nature of the effective action signifies the particle production in the system. Moreover the equation of the effective path $X(t)$ is determined by the variation of the effective action with respect to the external source ($J$) calculated at the vanishing of $J$ (see Eq. (\ref{5})). Hence the signature of particle production process must be encoded into the equation of the path through its complex nature. In fact, for the forced harmonic oscillator, the energy transferred to the system from the source was shown to be equal to imaginary part of the effective action which is in turn came out to be proportional to $|X(t)|^2$ in the large time limit (see the discussion in Section $3$ of \cite{ComplexPath1}). Therefore in \cite{ComplexPath1}, the rate of energy transferred is conjectured to de determined from variation of $|X(t)|^2$ with respect to time in the large time limit. Here we will now adopt the same conjecture and follow the identical steps, as done in \cite{ComplexPath1} to achieve the final goal.
The modulo square of $X(t)$ is then simply,
\begin{equation}
	\vert X(t)\vert^{2}=\dfrac{a_{0}}{\lambda}\Big\vert\dfrac{\cosh(\pi a_{0})}{\sinh(\pi a_{0})}\Big\vert~.
\end{equation}
According to the definition given in Eq. \eqref{eqn:different_symbols}, in the limit $t_{2}\rightarrow \infty$ and $ t_{1}=0$, $\lambda$ reduces to $m/2\hbar t$. Therefore, the modulo square of the effective path becomes,
\begin{equation}
	\vert X(t)\vert^{2}=\dfrac{2a_{0}\hbar t}{m}\Big(\dfrac{1}{e^{2\pi a_{0}}-1}+\dfrac{1}{2}\Big)~.
\end{equation}
Taking the time derivative, we get
\begin{equation}
	\dfrac{d\vert X(t)\vert^{2}}{dt}=\dfrac{2E}{m\kappa}\Big(\dfrac{1}{e^{\frac{2E\pi}{\hbar\kappa}}-1}+\dfrac{1}{2}\Big)~,
	\label{14}
\end{equation}
where $a_{0}=\dfrac{E}{\hbar\kappa}$ has been substituted.

Now, if one borrows the conjecture adopted in \cite{ComplexPath1}, the above quantity can be interpreted as the total energy transferred due to the interaction with the $1/x^2$ potential. Note that this one is in the form which is similar to the energy spectrum of harmonic oscillator $\sim N+1/2$ where the number of modes for the radiated wave $N$ is given by the Planck distribution. Therefore comparison with the standard form of Planck distribution, one identifies the temperature as $T=\hbar\kappa/2\pi$. Note that the temperature of the horizon takes the correct form. Moreover, since the starting Hamiltonian was obtained for SSSBH as well as Kerr black holes, here the complex path analysis has been successfully implemented both for static and stationary black holes. 

A point to be noted that the result has been obtained following the quantity $d|X(t)|^2/dt$ after interpreting it as the energy transfer by the source. As proposed in \cite{ComplexPath1}, such an interpretation is completely an analogy from the mathematical connection obtained for the forced harmonic oscillator. Till now there is no general proof, expect the example of the forced harmonic system.  
Furthermore Eq. (\ref{12}) has Fermionic nature, whereas, Eq. (\ref{14}) has Bosonic nature.  We already mentioned that the Hamiltonians, corresponding these two results, are not equivalent. Therefore it not expected that they will lead to identical results.  Moreover, the Hamiltonian $xp$ is completely relativistic in nature and has a resemblance with the nature of Dirac equation as it provides first order derivative with respect to space in wave equation. The later one may be the underlying reason for giving Fermionic nature in the distribution. Whereas, in the complex effective path approach, we are not directly investigating the $xp$ Hamiltonian. Rather we are studying a Hamiltonian which has the same wave function solution as that of $xp$ Hamiltonian and has a standard $p^{2}/2m+V(x)$ type form. This is a non-relativistic treatment and therefore the wave equation carries second order derivative with respect to position. Consequently the obtained result is Bosonic in nature. But notably the overall conclusion from these two completely disjoint quantum models is same -- the transition probabilities are thermal in nature and temperature of the horizon is that given by the Hawking expression. Finally, the structure of the Hamiltonian (\ref{eqn:near_horizon_hamiltonian}) was obtained through the near horizon expansion (see, \cite{Dalui:2018qqv,chaos2,Chaos3,Dalui:2021tvy} for details). In this case the relevant quantities have been expanded around the horizon $x=0$. Therefore to achieve this information about the location of the horizon is very much essential. However a similar Hamiltonian can be obtained near a generic null surface, particularly constructed in Gaussian null coordinates (see \cite{Dalui:2021sme}). But in this case $\kappa$ is no longer constant. In this sense such a structure of Hamiltonian may be generalised in the vicinity of any null surface, not necessarily it has to be an event horizon. On the other hand in calculating (\ref{14}) the path has been considered to be crossing the horizon at $x=0$. Apart from that information, the analysis has been done very near to the horizon. Therefore the role of structure of horizon seems to be very nominal in this analysis. 
 
\section{Conclusion}	
We implemented the complex path analysis (or path integral approach) of \cite{ComplexPath1} for SSSBH as well as Kerr spacetimes to find the Hawking temperature. The near horizon Hamiltonian of the form $\sim xp$ as was obtained in \cite{Dalui:2018qqv,chaos1,chaos2,Dalui:2021tvy,Chaos3} has been used to achieve our goal. We observed that the obtained value of temperature does not suffer from a factor of two issue. Therefore we feel that the adopted coordinate systems in the present case are very much suitable for studying Hawking radiation through the complex path. Note that the density of states as obtained in section \ref{DOS} (see Eq. (\ref{12})) is Fermionic in nature, whereas $N$ in (\ref{14}) is Bosonic one. The reason for this difference may be due to the two different ways of investigation. In DOS calculation, the Hamiltonian has been treated completely relativistically, whereas, in the complex path case, an effective potential has been obtained through the Schrodinger equation. Since the Schrodinger equation is valid in non-relativistic quantum mechanics, we feel that the obtained Hamiltonian has validity in the non-relativistic regime. Moreover, a relativistic theory boils down to the non-relativistic (Schrodinger) theory in the non-relativistic limit. Therefore such a difference in handling the model may be the cause for getting two types of distribution. Of course, it needs further investigation to provide a concrete reason. Nonetheless, in both ways, we obtained the correct value of Hawking temperature, which was our main goal.

It must be noted that our $xp$ Hamiltonian is unstable in nature. Hence, as proposed in \cite{ComplexPath1}, it might be the case that the unstable nature of the system may be liable for obtaining a complex nature of the path. Since such a Hamiltonian successfully describes a correct expression of horizon temperature in earlier approaches \cite{chaos2,Dalui:2021tvy,Chaos3,Dalui:2021sme,Chaos4} as well as through the present complex path approach, we again feel that the near horizon instability can be a core reason for thermalization of horizon. Therefore we elevate our earlier conjecture about the mechanism of the thermalization of the horizon as -- {\it the near horizon classical instability provides a complex path to a particle at the quantum level and thereby giving a finite probability to escape from the horizon, which is thermal in nature}. This observation again bolstered our earlier idea of explaining the thermal property of the horizon through local instability.

Few open issues, even in original work \cite{ComplexPath1}, and incompleteness exist in this complex path approach. Note that the required goal has been achieved from the quantity $d|X(t)|^2/dt$ by interpreting it as the energy transfer by the source. As we mentioned earlier, this is purely an analogy from the mathematical connection obtained for the forced harmonic oscillator. In fact, there is no general proof for this. Although this provides a correct temperature value of the horizon, a direct proof of such interpretation for the $1/x^2$ Hamiltonian is still lacking. Secondly, the present effective potential (\ref{eqn:potential}) is a complex function. Whether such complex nature has any deeper significance will be an important aspect to look at. Finally the energy of the ground state (Eq. (\ref{E8})) has been derived by subtracting a ``background'' contribution to make it meaningful and finite. It needs further investigation whether such a ``hand waving'' argument is practically correct. We aim to take up these issues in future.
	
\vskip 3mm
\noindent
{\bf Acknowledgment:}
GRK would like to express his gratitude to the Dept. of Physics at the Indian Institute of Technology, Guwahati, India, for giving the opportunity to do this research project and for supporting undergraduate research. Final part of the work was done when GRK was at Mathematical Institute, Oxford. The authors thank Suprit Singh for the extensive discussion and valuable comments.


\begin{widetext}
		
\section*{Appendices}
\appendix

\section{Alternative approach to DOS}\label{App3}
In order to evaluate (\ref{8}) we first transform Hamiltonian (\ref{eqn:near_horizon_hamiltonian}) to that of an inverted harmonic oscillator (IHO)($H=(\kappa/2)(P^2-X^2)$) by going into a new canonical variable
\begin{equation}
x=\frac{1}{\sqrt{2}}(P-X); \,\,\,\ p = \frac{1}{\sqrt{2}}(P+X)~.
\label{9}
\end{equation}
The propagator of the IHO can be obtained from that of harmonic oscillator by replacing frequency $\omega\to i\omega$. For (\ref{9}) frequency is identified as $\omega=\kappa$ and hence the propagator in this case comes out to be
\begin{eqnarray}
K(x_2=X,t_2=t; x_1=X,t_1=0) = \Big(\frac{i}{-2\pi\sinh(\kappa t)}\Big)^{1/2}\exp\Big[ibX^2\Big]~,
\label{10}
\end{eqnarray}
where $b=\frac{1}{\sinh(\kappa t)}\{\cosh(\kappa t) - 1\}$ (to find (\ref{10}) substitute $\omega = \kappa \to i\omega = i\kappa$ and $m=1/\kappa$ as well as $x_f=x_2=x_i=x_1=X, t_i=t_1=0, t_f=T=t_2=t$ in Eq. $3.81$ combined with Eq. $3.66$ of \cite{Book1}).
Then we have
\begin{eqnarray}
\int_{-\infty}^\infty dX~G(X,E) &=& \int_0^\infty dt e^{-\frac{i E t}{\hbar}} \int_0^\infty dx~K(x_2=X,t_2=t; x_1=X,t_1=0) 
\nonumber
\\
&=&\int_0^\infty dt~ \frac{e^{-\frac{i E t}{\hbar}}}{2\sinh(\frac{\kappa t}{2})}~.
\label{100}
\end{eqnarray}
In the above the $X$ integration has been done by replacing $b\to b+i\epsilon$ for $\epsilon>0$ and at the end $\epsilon\to 0$ has been taken.

The final $t$ integration can be done by replacing $t\to t-i\delta$ with $\delta>0$ and then choosing the contour in the fourth quadrant in complex $t$ plane. This will yield (\ref{12a}). 
The explicit steps are as follows. First change the integration variable as $t\to 2t/\kappa$. This yields
\begin{equation}
	\dfrac{1}{2}\int_{0}^{\infty}~dt~\dfrac{e^{-\frac{iEt}{\hbar}}}{\sinh(\frac{\kappa t}{2})}=\dfrac{1}{\kappa}\int_{0}^{\infty}~dt~\dfrac{e^{-i\frac{2 E}{\hbar\kappa}t}}{\sinh(t)}~.
\end{equation}
\begin{figure}[h!]
	\centering
	\includegraphics[width=6cm,height=6cm]{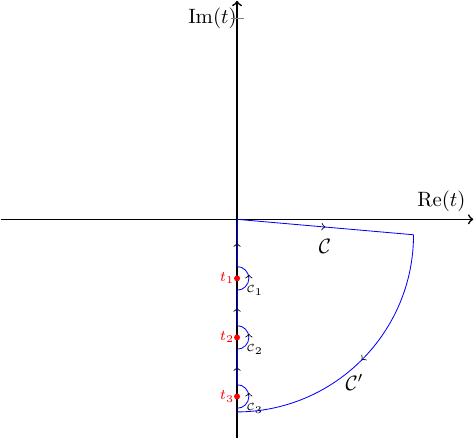}
	\caption{Contour plot for integration. The integration is done considering $t$ as a complex variable. The integration in the range $[0,\infty)$ is infinitesimally shifted from the real axis for the integral to be convergent.}
	\label{fig:contourplot}
\end{figure}\\
The integration is done using a suitable contour, as shown in Fig. \ref{fig:contourplot}. The motivation for choosing the contour can be found in \cite{Book2}. There are no poles inside the contour and hence, the total integral over the closed contour vanishes. However, the integration over the small semi-circles is non-zero. Hence, our required integral is negative of the sum of these integrals over the semi-circle: 
\begin{equation}
 \int_{\mathcal{C}}dt~f(t)+\int_{\mathcal{C}'}dt~ f(t)+\sum_{n=1}\int_{\mathcal{C}_{n}}dt ~f(t)+\int_{\mathcal{D}}dt~f(t)=0;~~f(t)=\dfrac{e^{-i\frac{2Et}{\kappa\hbar}}}{\sinh(t)}~.
\end{equation}
In the above $\mathcal{D}$ is the path on the imaginary axis between to consecutive poles.
If we look at the integration over a portion of the contour along the imaginary axis, it looks as follows
\begin{equation*}
	\int_{-i\pi+i\epsilon}^{0}\dfrac{e^{-i\frac{2Et}{\hbar\kappa}}}{\sinh(t)} dt=\int_{\pi-\epsilon}^{0}\dfrac{e^{-\frac{2Et}{\hbar\kappa}}}{\sin(t)} dt 
\end{equation*} 
This is a Real quantity. Hence, the integration over $\mathcal{D}$ is real, and won't contribute to the imaginary part. In this, the integral on the circular part, i.e, on the contour $\mathcal{C}'$ goes to zero as we take the limit of radius going to infinity as there is $\sinh(t)$ term in the denominator. We are interested in the imaginary part of the integration. This comes from the contribution made by integration over the infinitesimal semi-circles drawn over the poles. Using the Residue theorem, the integration on small semi-circular deformations will give us,
\begin{equation}
	\int_{\mathcal{C}_{n}}dt~f(t)=-\pi i\text{Res}(f(t_{n}));~~~t_{n}=-in\pi,~~n\in\mathbb{N}~.
\end{equation}
The negative sign is due to the direction in which the contour is traversed along. Note that $n=0$ is not included as for this we will have $t_0 = i\delta$ which is outside our chosen contour. Hence, we have
\begin{equation}
	\Im\Big(\int_{0}^{\infty}~dt~\dfrac{e^{-i\frac{2 E}{\hbar\kappa}t}}{\sinh(t)}\Big)=\pi \sum_{n=1}^{\infty}\text{Res}(f(t_{n}))~.
\end{equation}
The Residue at the poles are given as follows
\begin{equation}
	\text{Res}(f(t_{n}))=(-1)^{n}e^{-\frac{2En\pi}{\hbar\kappa}};~~t_{n}=-in\pi~.
\end{equation}
Hence, we have,
\begin{equation}
	\Im\Big(\dfrac{1}{2}\int_{0}^{\infty}~dt~\dfrac{e^{-\frac{iEt}{\hbar}}}{\sinh(\frac{\omega t}{2})}\Big)=\dfrac{\pi }{\omega}\sum_{n=1}^{\infty}(-e^{-\frac{2E\pi}{\hbar\kappa}})^{n}=-\dfrac{\pi }{\kappa}\dfrac{e^{-\frac{2\pi E}{\hbar\kappa}}}{1+e^{-\frac{2E\pi}{\hbar\kappa}}}=-\dfrac{\pi }{\kappa}\dfrac{1}{e^{\frac{2E\pi}{\hbar\kappa}}+1}
\end{equation}
and therefore DOS is given by (\ref{12}).

\section{Alternative way to effective potential} \label{App1}
The solution of \eqref{eqn:schrodinger1} provides the energy eigenfunction. This is given by
\begin{equation}
\psi_E(x)=C_0x^{\frac{iE}{\hbar\kappa}-\frac{1}{2}}~,
\label{eqn:wavefunction}
\end{equation}
where $C_0$ is the normalization constant. $C_0$ is determined through the nomalization condition $\int dx \psi_E'^*(x)\psi_E(x) = \delta(E-E')$ which yields $C_0=1/\sqrt{2\pi\hbar\kappa}$.

Now we ask the following question: {\it What is the effective potential under which a massive particle with mass $m$ will have energy eigen wavefunction, given by Eq. \eqref{eqn:wavefunction}, through the standard form of time-independent Schrodinger equation?}
In order to identify the effective potential, we have to substitute this wavefunction in the Schrodinger equation for a massive particle given by (\ref{B1}). Substitution of \eqref{eqn:wavefunction} in (\ref{B1}) yields the form of potential as
\begin{equation}
V(x)=\mathcal{E}+\dfrac{\hbar^{2}}{8mx^{2}}\Big(3-\dfrac{8iE}{\hbar\kappa}-\dfrac{4E^{2}}{\hbar^{2}\kappa^{2}}\Big)~.
\label{B2}
\end{equation}
This is the same as Eq. \eqref{eqn:potential} for a zero energy $(\mathcal{E})=0$ massive particle.

Just for the shake of completeness we provide a systematic steps to obtain the solution directly from the Schrodinger equation with the potential given by \eqref{eqn:potential}. Then the Hamiltonian of the system with a massive non-relativistic particle is given by, 
\begin{equation}
H=\dfrac{p^{2}}{2m}-\dfrac{k}{x^{2}};~~~k=\dfrac{\hbar^{2}}{8m}\Big(\dfrac{8iE}{\hbar\kappa}-3+\dfrac{4E^{2}}{\hbar^{2}\kappa^{2}}\Big)~.
\end{equation}
As the potential is time-independent, we can write $H\ket{\psi}=\mathcal{E}\ket{\psi}$.
Hence the Schrodinger equation is given by
\begin{equation}
-\dfrac{\hbar^{2}}{2m}\dfrac{d^{2}\psi}{dx^{2}}-\dfrac{k}{x^{2}}=\mathcal{E}\psi~.
\end{equation}
On rearranging the equation, we get,
\begin{equation}
x^{2}\psi''+\psi\Big[\dfrac{2mk}{\hbar^{2}}+\dfrac{2m\mathcal{E}}{\hbar^{2}}x^{2}\Big]=0~.
\label{eqn:general_schrodinger}
\end{equation}
Let us simplify the equation by expressing the wavefunction as $\psi(x)=C x^{1/2}f(x)$, where $C$ is a constant and $f(x)$ is a function to be determined. Substituting this in the \eqref{eqn:general_schrodinger} one finds 
\begin{equation}
x^{2}f''+xf'+f\Big[-\dfrac{1}{4}+\dfrac{2mk}{\hbar^{2}}+\dfrac{2m\mathcal{E}}{\hbar^{2}}x^{2}\Big]=0~.
\end{equation}
Changing the variable $\sqrt{\dfrac{2m\mathcal{E}}{\hbar^{2}}}x=y$,
and then substituting the value of $k$, we obtain
\begin{equation}
y^{2}\ddot{f}+y\dot{f}+f\Big[y^{2}-\Big(1-\dfrac{iE}{\hbar\kappa}\Big)^{2}\Big]=0~.
\end{equation}
This is the standard Bessel equation and hence the general solution is given by
\begin{equation}
\psi(x)=C x^{1/2}\Big(AJ_{\nu}\Big(\sqrt{\dfrac{2m\mathcal{E}}{\hbar^{2}}}x\Big)+BY_{\nu}(\sqrt{\dfrac{2m\mathcal{E}}{\hbar^{2}}}x\Big)\Big)~.
\label{eqn:general_solution}
\end{equation}
In the above we denoted $\nu=1-\dfrac{iE}{\hbar\kappa}$ and  $A,B$ are integration constants. 
We now proceed to study this solution under two limiting cases.

\subsection{Limiting Conditions}
\begin{enumerate}
	\item In the first limiting condition, we make the energy of the massive particle to be 0 i.e., $\mathcal{E}\rightarrow 0$. In this case, we know that the function must look like $x^{-1/2+\frac{iE}{\hbar\kappa}}$ (see Eq. \eqref{eqn:wavefunction}). The Limiting nature of the Bessel functions when $\vert z\vert\rightarrow 0$ are \cite{Besselfunctionproperties},
\begin{eqnarray}
&&J_{\nu}(z)\approx \dfrac{1}{\Gamma(\nu+1)}\Big(\dfrac{z}{2}\Big)^{\nu}~;
\nonumber
\\
&&Y_{\nu}(z)\approx -\dfrac{\Gamma(\nu)}{\pi}\Big(\dfrac{z}{2}\Big)^{-\nu}~.
\end{eqnarray}
Then under the limit $\mathcal{E}\to 0$, (\ref{eqn:general_solution}) yields	
\begin{equation}
\psi(x)\approx C x^{1/2}\Big[\dfrac{A}{\Gamma(\nu+1)}\Big(\sqrt{\dfrac{m\mathcal{E}}{2\hbar^{2}}}x\Big)^{1-\frac{iE}{\hbar\kappa}}-\dfrac{B\Gamma(1-\frac{iE}{\hbar\kappa})}{\pi}\Big(\sqrt{\dfrac{m\mathcal{E}}{2\hbar^{2}}}x\Big)^{-1+\frac{iE}{\hbar\kappa}}\Big]~.
\end{equation}
Note that the first term decays quickly as $\mathcal{E}\rightarrow 0$ and therefore will not contribute to this limiting case. Hence we have
\begin{equation}
\psi(x)\approx- C \dfrac{B\Gamma(1-\frac{iE}{\hbar\kappa})}{\pi}\Big(\sqrt{\dfrac{m\mathcal{E}}{2\hbar^{2}}}\Big)^{-1+\frac{iE}{\hbar\kappa}}x^{-\frac{1}{2}+\frac{iE}{\hbar\kappa}}~.
\end{equation}
Now in order to obtain the required form for $\mathcal{E}=0$ set $B=\Big(\sqrt{\dfrac{m\mathcal{E}}{2\hbar^{2}}}\Big)^{1-\frac{iE}{\hbar\kappa}}$ and then one obtains 
\begin{equation}
\psi(x)\approx -C \dfrac{\Gamma(1-\frac{iE}{\hbar\kappa})}{\pi}x^{-\frac{1}{2}+\frac{iE}{\hbar\kappa}}~.
\end{equation}
Finally use of normalization condition will yield \eqref{eqn:wavefunction}.	
	
\item Even though the investigation is done near the horizon, where we find the effective model having the form given by Eq.  \eqref{eqn:standard_hamiltonian}, we can look at the inverse-squared potential as a separate system and study the asymptotic limits, i.e. $x\rightarrow \infty$, or in other words, far away from the potential source. In this case, the particle should behave similarly to a free particle. The asymptotic values of the Bessel functions are \cite{Besselfunctionproperties}:
\begin{equation}
\begin{split}
J_{\nu}(z)\approx\sqrt{\dfrac{2}{\pi z}}\cos\Big(z-\nu\dfrac{\pi}{2}-\dfrac{\pi}{4}\Big)~;\\
Y_{\nu}(z)\approx\sqrt{\dfrac{2}{\pi z}}\sin\Big(z-\nu\dfrac{\pi}{2}-\dfrac{\pi}{4}\Big)~.
\end{split}
\end{equation}
Then (\ref{eqn:general_solution}) reduces to
\begin{equation}
\psi(x)\approx C\Big(\dfrac{2\hbar^{2}}{m\pi^{2}\mathcal{E}}\Big)^{1/4}\Big[A\cos\Big(\sqrt{\dfrac{2m\mathcal{E}}{\hbar^{2}}}x-\dfrac{3\pi}{4}+\dfrac{i\pi E}{2\hbar\kappa}\Big)
+B\sin\Big(\sqrt{\dfrac{2m\mathcal{E}}{\hbar^{2}}}x-\dfrac{3\pi}{4}+\dfrac{i\pi E}{2\hbar\kappa}\Big)\Big]~.
\end{equation}
This can be expanded as,
	\begin{equation}
		\begin{split}
			\psi(x)\approx C\Big(\dfrac{2\hbar^{2}}{m\pi^{2}\mathcal{E}}\Big)^{1/4}\Big[A\dfrac{e^{-\frac{\pi E}{2\hbar\kappa}}e^{i(\sqrt{\frac{2m\mathcal{E}}{\hbar^{2}}}x-\frac{3\pi}{4})}+e^{\frac{\pi E}{2\hbar\kappa}}e^{-i(\sqrt{\frac{2m\mathcal{E}}{\hbar^{2}}}x-\frac{3\pi}{4})}}{2}\\+B\dfrac{e^{-\frac{\pi E}{2\hbar\kappa}}e^{i(\sqrt{\frac{2m\mathcal{E}}{\hbar^{2}}}x-\frac{3\pi}{4})}-e^{\frac{\pi E}{2\hbar\kappa}}e^{-i(\sqrt{\frac{2m\mathcal{E}}{\hbar^{2}}}x-\frac{3\pi}{4})}}{2i}\Big]~.
		\end{split}
	\end{equation}
So both ingoing and outgoing free solutions are present.	
\end{enumerate}


\section{Evaluating the ratios $(I_{1}+I_{2})/D,~I_{3}/D$}\label{App2}
Before starting the calculations, we mention the integral identity used to simplify the equations\cite{integration_table}. 
\begin{equation}
\begin{split}
\int_{0}^{\infty} dx~x^{\lambda+1}e^{-\alpha x^{2}}J_{\mu}(\beta x)J_{\nu}(\gamma x)=\dfrac{\beta^{\mu}\gamma^{\nu}\alpha^{-(\mu+nu+\lambda+2)/2}}{2^{\mu+nu+1}\Gamma(\nu+1)}\sum_{m=0}^{\infty}\Big[\dfrac{\Gamma(m+\frac{1}{2}(\nu+\mu+\lambda+2))}{m!\Gamma(m+\mu+1)}\Big(\dfrac{-\beta^{2}}{4\alpha}\Big)^{m}\\
\times F(-m,-\mu-m;\nu+1;\dfrac{\gamma^{2}}{\beta^{2}})\Big]~.
\end{split}
\label{eqn:besselintegralindentity}
\end{equation}
The condition on $\mu,\nu,\lambda$ is $\text{Re}(\mu+\nu+\lambda)>-2$.
In the present case we identify $\mu=\nu=ia=1-ia_{0},~a_{0}=E/\hbar\kappa$ and $\lambda=1$. Hence the above mentioned condition is satisfied. We now evaluate our integrals. 
	
\subsection{Calculation of $(I_{1}+I_{2})/D$}
We start our calculations with evaluating the term $I_{1}$ (see Eq. (\ref{B3})). The integration of this expression to be evaluated is as follows:
\begin{eqnarray}
&&\int_{-\infty}^{\infty}dx~x^{2}e^{i\lambda x^{2}} J_{ia}(px)J_{ia}(qx)=\int_{-\infty}^{0}dx~x^{2}e^{i\lambda x^{2}} J_{ia}(px)J_{ia}(qx)+\int_{0}^{\infty}dx~x^{2}e^{i\lambda x^{2}} J_{ia}(px)J_{ia}(qx)
\nonumber
\\
&=& \int_{0}^{\infty}dx~x^{2}e^{i\lambda x^{2}} J_{ia}(-px)J_{ia}(-qx)+\int_{0}^{\infty}dx~x^{2}e^{i\lambda x^{2}} J_{ia}(px)J_{ia}(qx)
\nonumber
\\
&=& e^{2\pi a}\int_{0}^{\infty}dx~x^{2}e^{i\lambda x^{2}} J_{ia}(px)J_{ia}(qx)+\int_{0}^{\infty}dx~x^{2}e^{i\lambda x^{2}} J_{ia}(px)J_{ia}(qx)
\nonumber
\\
&=& (1+e^{2\pi a})\underbrace{\int_{0}^{\infty}dx~x^{2}e^{i\lambda x^{2}} J_{ia}(px)J_{ia}(qx)}_Y~.
\end{eqnarray}
Using the identity Eq. \eqref{eqn:besselintegralindentity} we find
\begin{equation}
\begin{split}
Y = \int_{0}^{\infty}dx~x^{2}e^{i\lambda x^{2}} J_{ia}(px)J_{ia}(qx)=\dfrac{p^{ia}q^{ia}(-i\lambda)^{-ia-3/2}}{2^{2ia+1}\Gamma(ia+1)}\sum_{m=0}^{\infty}\Big[\dfrac{\Gamma(m+ia+3/2)}{m!\Gamma(m+ia+1)}\Big(\dfrac{-p^{2}}{4i\lambda}\Big)^{m}\\
\times F(-m,-ia-m;ia+1;\dfrac{q^{2}}{p^{2}})\Big]~.
\end{split}
\end{equation}
Next substituting the values of $p,q,\lambda$ from Eq. \eqref{eqn:different_symbols} in  the above we obtain
\begin{equation}
\begin{split}
Y=\Big(\dfrac{2m\epsilon^{2}}{i\hbar(t_{2}-t_{1})}\Big)^{ia}\dfrac{(-i\lambda)^{-3/2}}{2^{2ia+1}\Gamma(ia+1)}\sum_{m=0}^{\infty}\Big[\dfrac{\Gamma(m+ia+3/2)}{m!\Gamma(m+ia+1)}\Big(\dfrac{-m\epsilon^{2}(t_{2}-t)}{2\hbar i(t_{2}-t_{1})(t-t_{1})}\Big)^{m}\\\times F\Big(-m,-ia-m;ia+1;\Big(\dfrac{t-t_{1}}{t_{2}-t}\Big)^{2}\Big)\Big]~.
\end{split}
\end{equation}
In the limit $\epsilon\rightarrow 0$, only $m=0$ term survives and then one obtains
\begin{equation}
Y=\Big(\dfrac{2m\epsilon^{2}}{i\hbar(t_{2}-t_{1})}\Big)^{ia}\dfrac{(-i\lambda)^{-3/2}}{2^{2ia+1}\Gamma(ia+1)}\dfrac{\Gamma(ia+3/2)}{\Gamma(ia+1)}\\\times F\Big(0,-ia;ia+1;\Big(\dfrac{t-t_{1}}{t_{2}-t}\Big)^{2}\Big)~.
\end{equation}
Later, we take the limit $t_{2}\rightarrow \infty$ in which case, the Hypergeometric function becomes $F(0,-ia;ia+1;0)=1$.
Hence we find 
\begin{equation}
I_{1}=\dfrac{e^{-2\pi a}(1+e^{2\pi a})}{\sinh^{2}(\pi a)}\Big(\dfrac{2m\epsilon^{2}}{i\hbar(t_{2}-t_{1})}\Big)^{ia}\dfrac{(-i\lambda)^{-3/2}}{2^{2ia+1}\Gamma(ia+1)}\dfrac{\Gamma(ia+3/2)}{\Gamma(ia+1)}~.
\end{equation}
Finally using the property $\Gamma(1-z)\Gamma(z)=\dfrac{\pi}{\sin\pi z}$ one finds
\begin{equation}
I_{1}=e^{\pi a/2}(1+e^{-2\pi a})\Big(\dfrac{m\epsilon^{2}}{2\hbar(t_{2}-t_{1})}\Big)^{ia}(-i\lambda)^{-3/2}\dfrac{[\Gamma(-ia)]^{2}\Gamma(ia+3/2)}{2\pi^{2}}~.
\end{equation}
In the integral $I_{2}$, there is $J_{-ia}$ instead of $J_{ia}$ (see Eq. (\ref{B4})). Hence, we can replace $a\rightarrow -a$ and get $I_{2}$
\begin{equation}
I_{2}=e^{-2\pi a}\times I_{1}(a\rightarrow -a)=e^{-\pi a/2}(1+e^{-2\pi a})\Big(\dfrac{m\epsilon^{2}}{2\hbar(t_{2}-t_{1})}\Big)^{-ia}(-i\lambda)^{-3/2}\dfrac{[\Gamma(ia)]^{2}\Gamma(-ia+3/2)}{2\pi^{2}}~.
\end{equation}

Till this point, the calculations are in agreement with the calculations done in \cite{ComplexPath1}. As mentioned earlier, we have,
\begin{equation}
a=-a_{0}-i;~~a_{0}=\dfrac{E}{\hbar\kappa}~.
\end{equation}
Then for our case	
\begin{equation}
\begin{split}
I_{1}= ie^{-\pi a_{0}/2}(1+e^{2\pi a_{0}})\dfrac{(-i\lambda)^{-3/2}}{2\pi^{2}}\exp\Big[\ln\Big(\dfrac{m\epsilon^{2}}{2\hbar(t_{2}-t_{1})}\Big)-ia_{0}\ln\Big(\dfrac{m\epsilon^{2}}{2\hbar(t_{2}-t_{1})}\Big)+i(2\theta_{1}+\theta_{2})\Big]\\\times \vert\Gamma(-1+ia_{0})\vert^{2}\vert\Gamma(5/2-ia_{0})\vert
\end{split}
\end{equation}
where, 
\begin{equation}
\theta_{1}=\arg(\Gamma(-1+ia_{0}));~~\theta_{2}=\arg(\Gamma(5/2-ia_{0}))
\end{equation}
and,
\begin{eqnarray}
&&\vert\Gamma(5/2-ia_{0})\vert=\Big\vert\dfrac{3}{2}-ia_{0}\Big\vert\Big\vert \dfrac{1}{2}-ia_{0}\Big\vert\vert\Gamma(1/2-ia_{0})\vert=\Big\vert\dfrac{3}{2}-ia_{0}\Big\vert\Big\vert \dfrac{1}{2}-ia_{0}\Big\vert\sqrt{\dfrac{\pi}{\cosh(\pi a_{0})}}~;
\\
\nonumber
&&\vert\Gamma(-1+ia_{0})\vert^{2}=\dfrac{\pi}{a_{0}\sinh(\pi a_{0})}\dfrac{1}{1+a_{0}^{2}}~.
\end{eqnarray}
Hence, using the above identities, we express $I_{1}$ as follows,
\begin{equation}
\begin{split}
I_{1}= ie^{-\pi a_{0}/2}(1+e^{2\pi a_{0}})\dfrac{(-i\lambda)^{-3/2}}{2\pi^{2}}\exp\Big[\ln\Big(\dfrac{m\epsilon^{2}}{2\hbar(t_{2}-t_{1})}\Big)-ia_{0}\ln\Big(\dfrac{m\epsilon^{2}}{2\hbar(t_{2}-t_{1})}\Big)+i(2\theta_{1}+\theta_{2})\Big]\\\times\Big\vert\dfrac{3}{2}-ia_{0}\Big\vert\Big\vert \dfrac{1}{2}-ia_{0}\Big\vert\sqrt{\dfrac{\pi}{\cosh(\pi a_{0})}}\times\dfrac{\pi}{a_{0}\sinh(\pi a_{0})}\dfrac{1}{1+a_{0}^{2}}~.
\end{split}
\end{equation}
Similarly, we can now express $I_{2}$ as
\begin{equation}
\begin{split}
I_{2}=-ie^{\pi a_{0}/2}(1+e^{2\pi a_{0}})\dfrac{(-i\lambda)^{-3/2}}{2\pi^{2}}\exp\Big[-\ln\Big(\dfrac{m\epsilon^{2}}{2\hbar(t_{2}-t_{1})}\Big)+ia_{0}\ln\Big(\dfrac{m\epsilon^{2}}{2\hbar(t_{2}-t_{1})}\Big)+i(2\theta_{3}+\theta_{4})\Big]\\\times \vert\Gamma(1-ia_{0})\vert^{2}\vert\Gamma(1/2+ia_{0})\vert~,
\end{split}
\end{equation}
where
\begin{equation}
\theta_{3}=\arg(\Gamma(1-ia_{0}));~~\theta_{4}=\arg(\Gamma(1/2+ia_{0})
\end{equation}
and, 
\begin{equation}
\vert\Gamma(1/2+ia_{0})\vert=\sqrt{\dfrac{\pi}{\cosh(\pi a_{0})}}~; \,\,\,\
\vert\Gamma(1-ia_{0})\vert^{2}=\dfrac{a_{0}\pi}{\sinh(\pi a_{0})}~.
\end{equation}
Hence we have
\begin{equation}
\begin{split}
I_{2}=-ie^{\pi a_{0}/2}(1+e^{2\pi a_{0}})\dfrac{(-i\lambda)^{-3/2}}{2\pi^{2}}\exp\Big[-\ln\Big(\dfrac{m\epsilon^{2}}{2\hbar(t_{2}-t_{1})}\Big)+ia_{0}\ln\Big(\dfrac{m\epsilon^{2}}{2\hbar(t_{2}-t_{1})}\Big)+i(2\theta_{3}+\theta_{4})\Big]\\\times \sqrt{\dfrac{\pi}{\cosh(\pi a_{0})}}\dfrac{a_{0}\pi}{\sinh(\pi a_{0})}~.
\end{split}
\end{equation}
Now that we have derived the expressions for $I_{1},I_{2}$, we add them:
\begin{equation}
\begin{split}
I_{1}+I_{2}=-ie^{\pi a_{0}/2}(1+e^{2\pi a_{0}})\dfrac{(-i\lambda)^{-3/2}}{2\pi^{2}}\exp\Big[-\ln\Big(\dfrac{m\epsilon^{2}}{2\hbar(t_{2}-t_{1})}\Big)+ia_{0}\ln\Big(\dfrac{m\epsilon^{2}}{2\hbar(t_{2}-t_{1})}\Big)+i(2\theta_{3}+\theta_{4})\Big]\\\times \sqrt{\dfrac{\pi}{\cosh(\pi a_{0})}}\dfrac{a_{0}\pi}{\sinh(\pi a_{0})}\Big\{-1+e^{-\pi a_{0}}  \exp\Big[2\ln\Big(\dfrac{m\epsilon^{2}}{2\hbar(t_{2}-t_{1})}\Big)-2ia_{0}\ln\Big(\dfrac{m\epsilon^{2}}{2\hbar(t_{2}-t_{1})}\Big)+i(2\theta_{1}+\theta_{2}\\-2\theta_{3}-\theta_{4})\Big]\times\Big\vert\dfrac{3}{2}-ia_{0}\Big\vert\Big\vert \dfrac{1}{2}-ia_{0}\Big\vert\dfrac{1}{a^{2}_{0}(1+a_{0}^{2})}\Big\}~.
\end{split}
\label{eqn:i1+i2}
\end{equation}
 In the limit $\epsilon\rightarrow 0$ there will be divergent terms in $I_{2}$. These will be cancelled when we take the ratio with $D$.

The denominator (given by (\ref{B6})), in the limiting case, is expressed as,
\begin{eqnarray}
D&=&-\dfrac{ie^{\pi a_{0}}}{\pi}\Big(-e^{-\pi a_{0}}\Gamma(-1+ia_{0})\Big(\dfrac{m\epsilon^{2}}{2\hbar(t_{2}-t_{1})}\Big)^{ia}+\Gamma(1-ia_{0})\Big(\dfrac{m\epsilon^{2}}{2\hbar(t_{2}-t_{1})}\Big)^{-ia}\Big)
\\
\nonumber
&=&-\dfrac{ie^{\pi a_{0}}}{\pi}\Big(-e^{-\pi a_{0}}\sqrt{\dfrac{\pi}{a_{0}\sinh(\pi a_{0})}\dfrac{1}{1+a_{0}^{2}}}\exp\Big[\Big(\ln\dfrac{m\epsilon^{2}}{2\hbar(t_{2}-t_{1})}\Big)-ia_{0}\ln\Big(\dfrac{m\epsilon^{2}}{2\hbar(t_{2}-t_{1})}\Big)+i\theta_{1}\Big]
\nonumber
\\
&+&\sqrt{\dfrac{a_{0}\pi}{\sinh(\pi a_{0})}}\exp\Big[-\ln\Big(\dfrac{m\epsilon^{2}}{2\hbar(t_{2}-t_{1})}\Big)+ia_{0}\ln\Big(\dfrac{m\epsilon^{2}}{2\hbar(t_{2}-t_{1})}\Big)+i\theta_{3}\Big]\Big)
\nonumber
\\
&=& -\dfrac{ie^{\pi a_{0}}}{\pi}\sqrt{\dfrac{a_{0}\pi}{\sinh(\pi a_{0})}}\exp\Big[-\ln\Big(\dfrac{m\epsilon^{2}}{2\hbar(t_{2}-t_{1})}\Big)+ia_{0}\ln\Big(\dfrac{m\epsilon^{2}}{2\hbar(t_{2}-t_{1})}\Big)+i\theta_{3}\Big] 
\nonumber
\\
&\times& \Big(-e^{-\pi a_{0}}\dfrac{1}{a_{0}}\sqrt{\dfrac{1}{1+a_{0}^{2}}}\exp\Big[2\ln\Big(\dfrac{m\epsilon^{2}}{2\hbar(t_{2}-t_{1})}\Big)-2ia_{0}\ln\Big(\dfrac{m\epsilon^{2}}{2\hbar(t_{2}-t_{1})}\Big)+i(\theta_{1}-\theta_{3})\Big]+1\Big)~.
\label{eqn:D}
\end{eqnarray}
Now, upon taking the ratio of Eq. \eqref{eqn:D} and Eq. \eqref{eqn:i1+i2}, we get,
\begin{equation}
\dfrac{I_{1}+I_{2}}{D}=e^{-\pi a_{0}/2}(1+e^{2\pi a_{0}})\dfrac{(-i\lambda)^{-3/2}}{2}\sqrt{\dfrac{a_{0}}{\cosh(\pi a_{0})\sinh(\pi a_{0})}}e^{i(\theta_{3}+\theta_{4})}\times \zeta
\label{B7}
\end{equation}
where,
\begin{equation}
\zeta=\dfrac{-1+e^{-\pi a_{0}}  \exp\Big[2\ln\Big(\dfrac{m\epsilon^{2}}{2\hbar(t_{2}-t_{1})}\Big)-2ia_{0}\ln\Big(\dfrac{m\epsilon^{2}}{2\hbar(t_{2}-t_{1})}\Big)+i(2\theta_{1}+\theta_{2}\\-2\theta_{3}-\theta_{4})\Big]\dfrac{\vert\frac{3}{2}-ia_{0}\vert\vert \frac{1}{2}-ia_{0}\vert}{a^{2}_{0}(1+a_{0}^{2})}}{-e^{-\pi a_{0}}\dfrac{1}{a_{0}}\sqrt{\dfrac{1}{1+a_{0}^{2}}}\exp\Big[2\ln\Big(\dfrac{m\epsilon^{2}}{2\hbar(t_{2}-t_{1})}\Big)-2ia_{0}\ln\Big(\dfrac{m\epsilon^{2}}{2\hbar(t_{2}-t_{1})}\Big)+i(\theta_{1}-\theta_{3})\Big]+1}~.
\end{equation}
In the limit $\epsilon \rightarrow 0$ we have $\exp\Big[\ln\Big(\dfrac{m\epsilon^{2}}{2\hbar(t_{2}-t_{1})}\Big)\Big]\rightarrow 0$ and hence, 
$\lim_{\epsilon\rightarrow 0}\zeta=-1$. Then (\ref{B7}) reduces to the simple form Eq. (\ref{B8}).

\subsection{Evaluation of $I_{3}/D$}
$I_3$ is given by (\ref{B5}).
Use of the identity Eq. \eqref{eqn:besselintegralindentity} yields
\begin{equation}
I_{3}=-\dfrac{2e^{-\pi a}}{\sinh^{2}(\pi a)}\Big[\dfrac{(p/q)^{ia}(-i\lambda)^{-3/2}}{2\Gamma(1-ia)}\sum_{m=0}^{\infty}\dfrac{\Gamma(m+\frac{3}{2})}{m!\Gamma(m+ia+1)}\Big(\dfrac{p^{2}}{4i\lambda}\Big)^{m}F\Big(-m,-ia-m;-ia+1;\dfrac{q^{2}}{p^{2}}\Big)+(a\rightarrow -a)\Big]~.
\end{equation}
In the above expression, $(a\rightarrow -a)$ means, replace ``$a$" in the terms before positive sign which are inside the third bracket by ``$- a$". 
Under the limits the above reduces to
\begin{equation}
	I_{3}=-\dfrac{2e^{-\pi a}}{\sinh^{2}(\pi a)}\Big[\dfrac{(-i\lambda)^{-3/2}\Gamma(3/2)}{2\Gamma(1-ia)\Gamma(ia+1)}\Big(-\dfrac{t_{2}-t}{t-t_{1}}\Big)^{ia}+\dfrac{(-i\lambda)^{-3/2}\Gamma(3/2)}{2\Gamma(1+ia)\Gamma(1-ia)}\Big(-\dfrac{t_{2}-t}{t-t_{1}}\Big)^{-ia}\Big]
	\end{equation}
Note that $I_{3}$ independent of $\epsilon$. Then it is clearly visible that, due to the divergent part for the limit $\epsilon\to0$ in $D$ (see Eq. \eqref{eqn:D}),  $\lim_{\epsilon\rightarrow 0} (1/D)=0$. Hence, in the limit $\epsilon\rightarrow0$, $I_{3}/D$ vanishes and so we have (\ref{B9}).

\vskip 3mm
\noindent
{\Large{\bf{Declarations}}}
\vskip 2mm
\noindent
{\bf{Ethical Approval:}} Not applicable
\vskip 2mm
\noindent
{\bf Competing interests:} Not applicable
\vskip 2mm
\noindent
{\bf Authors' contributions:} Both the authors equally contributed right from the calculations and interpreting the results.
\vskip 2mm
\noindent
{\bf Funding:} There is no funding for this work to mention.
\vskip 2mm
\noindent
{\bf Availability of data and materials:} No data was used or produced in this work.		
\end{widetext}

	

\end{document}